\documentclass[sigconf]{acmart}

\usepackage{multirow}

\copyrightyear{2026}
\acmYear{2026}
\setcopyright{cc}
\setcctype{by}
\acmConference[ICAIF '26]{7th ACM International Conference on AI in Finance}{November 14--17, 2026}{Milan, Italy}
\acmBooktitle{7th ACM International Conference on AI in Finance (ICAIF '26), November 14--17, 2026, Milan, Italy}
\acmDOI{10.1145/3858063.3858862}
\acmISBN{979-8-4007-3087-0/2026/11}

\begin{document}

\author{Valentin Mohl}
\orcid{0000-0002-6166-2222}
\authornote{Equal contribution.}
\email{valentin.mohl@cs.ox.ac.uk}
\affiliation{%
  \department{British Open-ended Learning and Discovery Lab (BOLD)}
  \department{Department of Computer Science}
  \institution{University of Oxford}
  \city{Oxford}
  \country{United Kingdom}
}
\author{Alexandre Bismuth}
\orcid{0009-0008-6558-2781}
\authornotemark[1]
\email{abismuth@g.harvard.edu}
\affiliation{%
  \department{British Open-ended Learning and Discovery Lab (BOLD)}
  \department{Department of Computer Science}
  \institution{University of Oxford}
  \city{Oxford}
  \country{United Kingdom}
}
\affiliation{%
  \institution{Harvard University}
  \city{Cambridge}
  \state{MA}
  \country{United States}
}
\author{Aramis Marti-Shahandeh}
\orcid{0009-0007-4757-4844}
\email{aramis@robots.ox.ac.uk}
\affiliation{%
  \department{British Open-ended Learning and Discovery Lab (BOLD)}
  \department{Department of Engineering Science}
  \institution{University of Oxford}
  \city{Oxford}
  \country{United Kingdom}
}
\author{George Nigmatulin}
\orcid{0000-0001-8935-1146}
\email{george.nigmatulin@eng.ox.ac.uk}
\affiliation{%
  \department{British Open-ended Learning and Discovery Lab (BOLD)}
  \department{Department of Engineering Science}
  \institution{University of Oxford}
  \city{Oxford}
  \country{United Kingdom}
}

\author{Kang Li}
\orcid{0009-0007-7847-9292}
\email{kang.li@stats.ox.ac.uk}
\affiliation{%
  \department{British Open-ended Learning and Discovery Lab (BOLD)}
  \department{Department of Statistics}
  \institution{University of Oxford}
  \city{Oxford}
  \country{United Kingdom}
}
\author{Stefan Zohren}
\orcid{0000-0002-3392-0394}
\email{stefan.zohren@eng.ox.ac.uk}
\affiliation{%
  \department{British Open-ended Learning and Discovery Lab (BOLD)}
  \department{Department of Engineering Science}
  \institution{University of Oxford}
  \city{Oxford}
  \country{United Kingdom}
}
\author{Jakob Foerster}
\orcid{0000-0001-9688-2498}
\email{jakob@robots.ox.ac.uk}
\affiliation{%
  \department{British Open-ended Learning and Discovery Lab (BOLD)}
  \department{Department of Engineering Science}
  \institution{University of Oxford}
  \city{Oxford}
  \country{United Kingdom}
}
\author{Mihai Cucuringu}
\orcid{0000-0002-8464-2152}
\email{mihai@math.ucla.edu}
\affiliation{%
  \department{Department of Mathematics}
  \institution{University of California, Los Angeles}
  \city{Los Angeles}
  \country{United States}
}
\affiliation{%
  \department{Oxford-Man Institute of Quantitative Finance, Department of Statistics}
  \institution{University of Oxford}
  \city{Oxford}
  \country{United Kingdom}
}
\author{Anisoara Calinescu}
\orcid{0000-0003-2082-734X}
\email{ani.calinescu@cs.ox.ac.uk}
\affiliation{%
  \department{British Open-ended Learning and Discovery Lab (BOLD)}
  \department{Department of Computer Science}
  \institution{University of Oxford}
  \city{Oxford}
  \country{United Kingdom}
}

\title{Understanding the Inner Workings of a State-Space Foundation Model for Limit Order Books}

\begin{abstract} 
Autoregressive foundation models offer state-of-the-art capabilities to simulate the market by generating the message flow of limit order books (LOBs). Yet, without understanding their inner workings, research directions are hard to narrow and practical use in real-world trading environments remains uncertain. This study therefore asks what these models actually learn about market microstructure. We find that, similarly to games where sequence models build an internal representation of the board, an 8.2M-parameter Mamba-3 state-space model trained autoregressively on the tokenized message stream of eight large-cap US equities learns an internal representation of the market and its dynamics. In particular, order-flow imbalance, a central driver of short-horizon price moves, is decodable at $R^2{\approx}0.59$, far above controls and baselines. This representation extends to unseen stocks, suggesting that these models learn general representations rather than ticker-specific patterns. The model also generalizes across metrics beyond order-flow imbalance: results hold across traded volume and volatility. Additionally, the representation is causally steerable: pushing the state along a feature's axis shifts the model's next-message beliefs as that feature implies, monotonically in dose, consistent with the emergence of a world model.
\end{abstract}



\maketitle

\renewcommand{\shortauthors}{Mohl, Bismuth et al.}

\section{Introduction}

Financial markets produce sequence data at a scale few domains can match. The message streams of the S\&P~500 constituents alone comprise trillions of tokens of nanosecond-resolution order flow, similar to the size of text corpora used to train frontier language models~\cite{grattafiori2024llama}. This abundance is beginning to power a new class of models: autoregressive foundation models trained directly on tokenized limit order book (LOB) messages, which generate realistic order flow~\cite{NagyGenerativeAI2023,wheeler2024marketgpt,li2025mars}, and outperform traditional model classes such as GANs and parametric simulators on distributional benchmarks~\cite{nagy2025lobbench}. Yet, these models are black boxes. Backtested performance can justify deployment on the distributions it was measured on, but says nothing about behavior when the market shifts, and thus cannot earn trust in tail events, where mistakes are costliest. Without access to what a model represents, it also becomes harder to decide where to improve it. Understanding LOB foundation models from the inside is therefore of practical as well as scientific value.
 
What such a model learns is a scientific question with a well-known template: sequence models trained only on Othello moves spontaneously develop an internal representation of the board~\cite{li2023emergent,nanda2023emergentlinear}, and similar emergent world representations arise in chess and program execution~\cite{karvonen2024chess,jin2024karel}. We thus investigate whether a model trained only on the message stream builds an internal representation of the market, whose order book it is never shown. To study this question, we consider an 8.2M-parameter Mamba-3 state-space model (SSM)~\cite{lahoti2026mamba} trained autoregressively on four years of data from eight large-cap US equities, an efficient model that scores competitively on LOB-Bench~\cite{nagy2025lobbench}. 

We find that the market representation lives in the recurrent state rather than the residual stream. The state maintains the market's \emph{dynamics}: realized order-flow imbalance (OFI), a central driver of short-horizon price moves in the microstructure literature~\cite{cont2014price}, decodes at $R^2{\approx}0.59$ against ${\approx}0$ for a random-weights twin, with traded volume and realized volatility alongside. Because Mamba-3's recurrence admits an exact attention form~\cite{dao2024mamba2,lahoti2026mamba}, we can attribute the state message by message: the flow signal is integrated by a small set of learned long-memory heads with half-lives of 28--66 messages. The representation generalizes across stocks, including ones never seen in training. Additionally, it is causally steerable: pushing the state along the OFI direction shifts the model's next-message expectations monotonically in dose, against matched null directions. Its forward information is limited, however: a linear readout of the frozen state predicts the next 25 messages' mid-price move at only $R^2{\approx}0.04$, on par with a readout of the visible book.

In summary, our contributions are threefold:
\begin{enumerate}
  \item We present, to our knowledge, the first mechanistic interpretability study of a foundation model trained natively on limit-order-book data, extending emergent-world-model analysis to real market data, a noisy, partially observed, high-stakes domain.
  \item We characterize what is represented: not just a picture of the market but its flow, which generalizes across stocks, including ones never seen in training.
  \item We show the representation is causally steerable: steering the state shifts the model's beliefs monotonically in dose against matched-norm nulls.
\end{enumerate}

\section{Background}

A limit order book (LOB) collects the outstanding buy (bid) and sell (ask) limit orders of a stock at different price levels~\cite{gould2013limit,bouchaud2018trades}, with the best bid and ask (the \emph{touch}) separated by the spread and orders queuing under price-time priority. All activity arrives as a stream of messages at nanosecond resolution: submissions, cancellations, and executions. Each message is consumed by our model as a 26-token block that extends the tokenization of  \citet{NagyGenerativeAI2023}.

From this stream one can compute the market statistics we probe for, most importantly order-flow imbalance (OFI)~\cite{cont2014price}, a standard short-horizon predictor in the microstructure literature~\cite{kolm2023deep}. OFI measures the net buy-side pressure over a lookback window of $W$ messages:

\begin{equation}
    \begin{aligned}
    e^{b}_n &= \mathbf{1}\{P^{b}_n \ge P^{b}_{n-1}\}\, q^{b}_n
             - \mathbf{1}\{P^{b}_n \le P^{b}_{n-1}\}\, q^{b}_{n-1}, \\
    e^{a}_n &= \mathbf{1}\{P^{a}_n \le P^{a}_{n-1}\}\, q^{a}_n
             - \mathbf{1}\{P^{a}_n \ge P^{a}_{n-1}\}\, q^{a}_{n-1}, \\
    \mathrm{OFI}(W) &= \sum_{n \in W} \bigl(e^{b}_n - e^{a}_n\bigr),
    \end{aligned}
\end{equation}

where $P_n$ and $q_n$ are the best prices and queue sizes after event $n$ (one event per message). We follow \citet{cont2014price} and compute OFI from the reconstructed book's best-quote prices and aggregate queue sizes after each message. Positive values correspond to net buying pressure, while negative values indicate net selling pressure. For depth $k$, following \citet{xu2018multi}, let $e^{b,l}_n$ and $e^{a,l}_n$ denote the same expressions evaluated at the $l$-th best bid and ask, respectively; then $\mathrm{OFI}_k(W) = \sum_{n \in W} \sum_{l=1}^{k} \bigl(e^{b,l}_n - e^{a,l}_n\bigr)$, with $\mathrm{OFI}(W) = \mathrm{OFI}_1(W)$ the standard case above. 

\section{Related Work}

Sequence models trained on next-token prediction can develop internal representations of the hidden state underlying their training data: a Transformer trained only on Othello moves builds an internal board representation~\cite{li2023emergent}, later shown to be linear in the right basis~\cite{nanda2023emergentlinear}, with similar findings in chess~\cite{karvonen2024chess} and program execution~\cite{jin2024karel}, and human-interpretable concepts even in board-observing AlphaZero~\cite{mcgrath2022chess}. Probe accuracy and predictive success can nonetheless overstate the coherence of the underlying world model~\cite{vafa2024evaluating,vafa2025inductive}, and probed states can be causally inert~\cite{balogh2026adversarial}, motivating our interventional tests. Beyond games, sequence models encode static real-world properties such as protein structure~\cite{rives2021protein} and spatial and temporal coordinates~\cite{gurnee2024space}. For SSMs specifically, attribution methods recast or decompose the token-level computation~\cite{ali2024hidden,jafari2024mambalrp,pitorro2025latim}, and steering has so far been explored only in language models~\cite{paulo2024rnns,young2026writesae,sunkumohan2026bottleneck}. Whether an emergent representation arises in a model of a noisy, real-world dynamical system such as a financial market has remained open.

In the financial domain, deep models for market microstructure range from supervised LOB predictors~\cite{zhang2019deeplob,wallbridge2020translob,sirignano2019universal,prata2024lobcast,jung2025attention} to GAN-, RNN-, and diffusion-based generators~\cite{coletta2022learning,cont2023lobgan,hultin2023generative,berti2025trades,backhouse2025painting,wang2026difflob}. Recently, foundation models trained directly on the message stream have emerged~\cite{NagyGenerativeAI2023,wheeler2024marketgpt,li2025bytegen,linna2025lobert,kawawabeaudan2026tradefmgenerativefoundationmodel,li2025mars}, with the autoregressive approach outperforming GAN-based and parametric alternatives on distributional benchmarks~\cite{nagy2025lobbench}. The evaluation of all these models, however, has remained largely external, on predictive accuracy and distributional realism rather than on what they compute internally.

The interpretability of deep market models has so far mostly relied on post-hoc attribution, which indicates \emph{which} inputs influence a prediction but not \emph{how} they are processed~\cite{lundberg2017unified,ribeiro2016why}: sensitivity analyses of DeepLOB~\cite{zhang2019deeplob} and robustness analyses of generative LOB models~\cite{coletta2023explainability}. Mechanistic analysis in finance has so far largely targeted language models applied to financial text~\cite{golgoon2024mechanistic,tatsat2025beyond,molinari2025interpretable}, including concept steering~\cite{chen2025financial}, while interventions on time-series foundation models steer generic signal properties~\cite{wilinski2025tsfm} or transplant bulk activation statistics between market regimes~\cite{sanyal2025time2time}. Closest in spirit to our study, concurrent work probes the latent volatility state of a Transformer trained on simulated stochastic-volatility returns~\cite{huang2026emergent}, not a foundation model of real order-book messages. To our knowledge, no prior work applies mechanistic interpretability to a foundation model trained natively on limit-order-book data, nor probes or steers an SSM's recurrent state outside of language and synthetic game domains.

\section{Method}
\label{sec:method}

\subsection{Training pipeline} 

Our study leverages a state-space model using the Mamba-3 architecture~\cite{lahoti2026mamba}. We select an efficient variant with 8.2M parameters, a residual-stream width of $d_{\text{model}}=256$, and 13 SSM layers, trained autoregressively on the tokenized message stream. To make probing for book statistics meaningful, the model receives only the message stream and is never given the order book as input. Despite its small parameter count and the absence of any book input, this model scores competitively on LOB-Bench~\cite{nagy2025lobbench}, with a score close to larger models while having a faster inference speed.

Each SSM block maintains a fixed-size recurrent state, updated per token. Per head, the state is a $64\times128$ matrix; with 8 heads, this gives $65{,}536$ state values per layer.

We use LOBSTER-format \cite{huang2011lobster} limit order book message streams (from which we remove the book) to train our model on eight large-cap US equities (GOOG, AAPL, NVDA, AMZN, META, TSLA, MSFT, AMD) over the 2022-01 to 2025-12 period with windows of 500 messages each. With 26 tokens per message, each 500-message window spans $13{,}000$ tokens. For mechanistic interpretability, probing tests performance on GOOG stock over January 2026, a period held out during training. Generalization experiments also test on the other seven training tickers over the same month as well as on JPM and ORCL, two out-of-distribution stocks. We evaluate our model over $N=2{,}014$ windows spanning 20 trading days.

\subsection{Probing}
\label{sec:probing}

\begin{figure*}[tb]
    \centering
    \includegraphics[width=\textwidth]{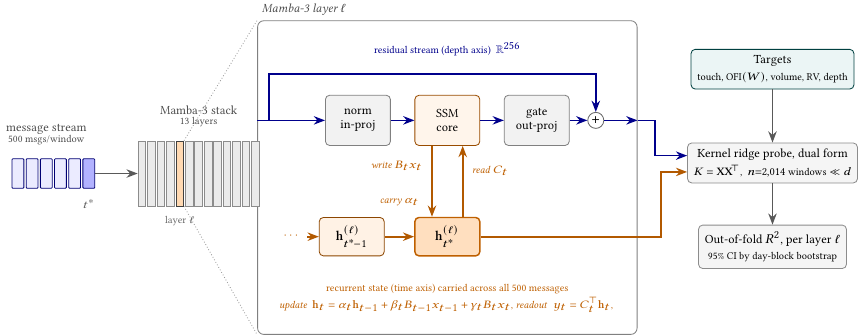}
    \Description{Diagram of a Mamba-3 block showing where the residual stream and the recurrent state are probed.}
    \caption{Probing loci in a message-only Mamba-3 model. At each token, the state carries over the previous state scaled by $\alpha_t$ and adds two writes: the current token's, weighted by $\gamma_t$, and the previous token's, weighted by $\beta_t$. From each layer we extract the per-token residual stream and the recurrent state $\mathbf{h}^{(\ell)}_t$, which accumulates across all messages of the window. Both are read out by the same kernel ridge probe.}
    \label{fig:probing}
\end{figure*}

We probe various market metrics in (a) the per-layer residual stream ($256$-dim) and (b) the per-layer recurrent state, flattened to a 65,536-dimensional vector, each at a fixed message boundary near the end of the window (Figure~\ref{fig:probing}). The targets are the touch (half the spread, in ticks), OFI over the window (Section~2), traded volume (the log total size of visible executions over the window), realized volatility (the standard deviation of per-message mid-price changes over the window), and book depth (the log total resting volume on the ten best levels of both sides, a proxy for available liquidity), all at the observation boundary.

Beyond what the state contains, we quantify each past message's contribution to the recurrent state. The Mamba-3 recurrence has an equivalent attention form~\cite{dao2024mamba2,lahoti2026mamba}, so the recurrent state at the boundary is exactly a weighted sum of one write per token. Summing the writes of a message's 26 tokens gives that message's contribution. We recover those weights from the model's own gates and verify on every window that they reproduce its state and layer output up to accumulated floating-point rounding. For each message we then record how much it writes into the state, and how much of what the OFI decoder reads out comes from it.

\subsection{Causal steering}
\label{sec:steering-method}

A probe can establish only that a quantity is \emph{present} in the state, not that the model \emph{uses} it~\cite{belinkov2022probing}. The test for use is interventional: change the internal representation, and check that the output changes in the way the representation's meaning dictates. Causal steering, adapted from activation steering in language models~\cite{turner2023activation,rimsky2024steering}, writes a belief into the recurrent state and reads out its consequences. The direction we write is, for each steered feature (OFI, traded volume, realized volatility), the mean recurrent state of that feature's top-tercile windows minus that of its bottom-tercile ones, added in all 13 layers to the state the model has built from the window, at the observation boundary. One dose ($\alpha{=}1$) is exactly that full high--low step (for OFI, from sell- to buy-side pressure). The dose grid is seven equally spaced doses from the 1st to the 99th percentile of where real windows sit along the direction, as well as $\alpha{=}0$. Effects are judged, at every dose, against 30 shuffled-label axes at matched per-layer norm and against an isotropic random direction at the same norm. 

The primary readout is the model's own next-token distribution. After the injection we teacher-force the real future messages and record, for the next message, the belief each feature should move: the probability of a buy for order flow, the probability of an execution for traded volume, and the perplexity of the price forecast for realized volatility. Each is compared with an unperturbed pass over the identical window, so there is no sampling noise and window-specific market conditions cancel. 

A second readout reaches the market itself: the model generates future messages freely, which are then replayed through the matching engine. In the market, an order-flow imbalance moves the price through the orders themselves: they arrive at the book, consume liquidity, and shift the quotes. Neither readout reproduces that process in full. The next-token readout holds future messages fixed by teacher forcing: the executions never happen, and what we measure is the shift in the model's expectation of them. The engine replay lets the model generate its own messages instead, so executions do occur, but no real orders arrive and the price moves only through what the model chooses to generate. The effect is indirect, and it carries the sampling and replay variance that the paired comparison removes from the next-token readout. We thus treat the next-token distribution as the primary causal measurement and the engine replay as a secondary check (Section~\ref{sec:steering}).

\subsection{Forward predictions}
\label{sec:forward-method}

Beyond what the state encodes about the present and how the model uses it, we also investigate whether this information is useful to predict the future. We ask this in two ways, reading from the frozen state: a regression of future market statistics, and a classification of the direction of the future mid-price move. Future-time decoding uses the same probe on longer windows: 700 messages tapped at the same boundary, so every window carries at least 200 real future messages after the boundary. For the regression, the targets are the realized mid-price move and the realized level-1 and top-5 OFI over the next $H$ messages, taken from the real data after the boundary, not from a model rollout, so the state is read once and the model never generates. 

To evaluate directional accuracy, we fit a one-hot kernel-ridge classifier with the same cross-validation, on windows with 1{,}500 messages so that at least 1{,}000 real future messages follow the boundary: three classes (down, flat, up), with flat the windows whose mid does not move, and a binary up-versus-down variant restricted to the windows that do. We report accuracy and macro-F1 against the majority-class baseline, since the class balance shifts markedly with $H$.

\subsection{Robustness}

Evaluating market decodability robustly requires careful consideration. Given that features ($65{,}536$) far exceed samples ($2{,}014$), our probes are solved as a kernel (dual-form) ridge regression with an $n\times n$ Gram system. Even though features $\gg$ samples, the OFI probe's effective degrees of freedom are only $\approx1\%$ of the state dimensions under ridge penalty. A PCA of 64 components recovering full $R^2$ performance also confirms the absence of overfitting. Then, held-out $R^2$ is computed under \emph{day-grouped} cross-validation, with the layer and $\lambda$ chosen by inner CV on the training days. Given time-series-sensitive targets we also compare our results with \emph{forward-chaining} (expanding-window) cross-validation. 95\% confidence intervals are then computed via day-block bootstrap to mitigate estimator variance. 

On top of that, we use two complementary controls. (i) A random-weights twin with identical architecture separates learned representations from architectural memory or random feature expansion. (ii) Book-snapshot baselines measure what is recoverable directly from the book: linear and RBF ridge readouts of 28 snapshot features of the L10 book (price offsets, spread, imbalances, microprice, per-level depths) taken at the window start and at the observation boundary, and of the pair augmented with the window's mid move and duration. Since the model never receives the book, whatever its state encodes was built from messages alone, and decodability above a single-snapshot baseline shows that the state holds information a snapshot of the current book does not carry.

\section{Results}

\subsection{Encoding of market statistics}

\subsubsection{Residual stream vs. recurrent state}
\label{sec:residual_vs_state}

\begin{figure}[h]
    \centering
    \includegraphics[width=\columnwidth]{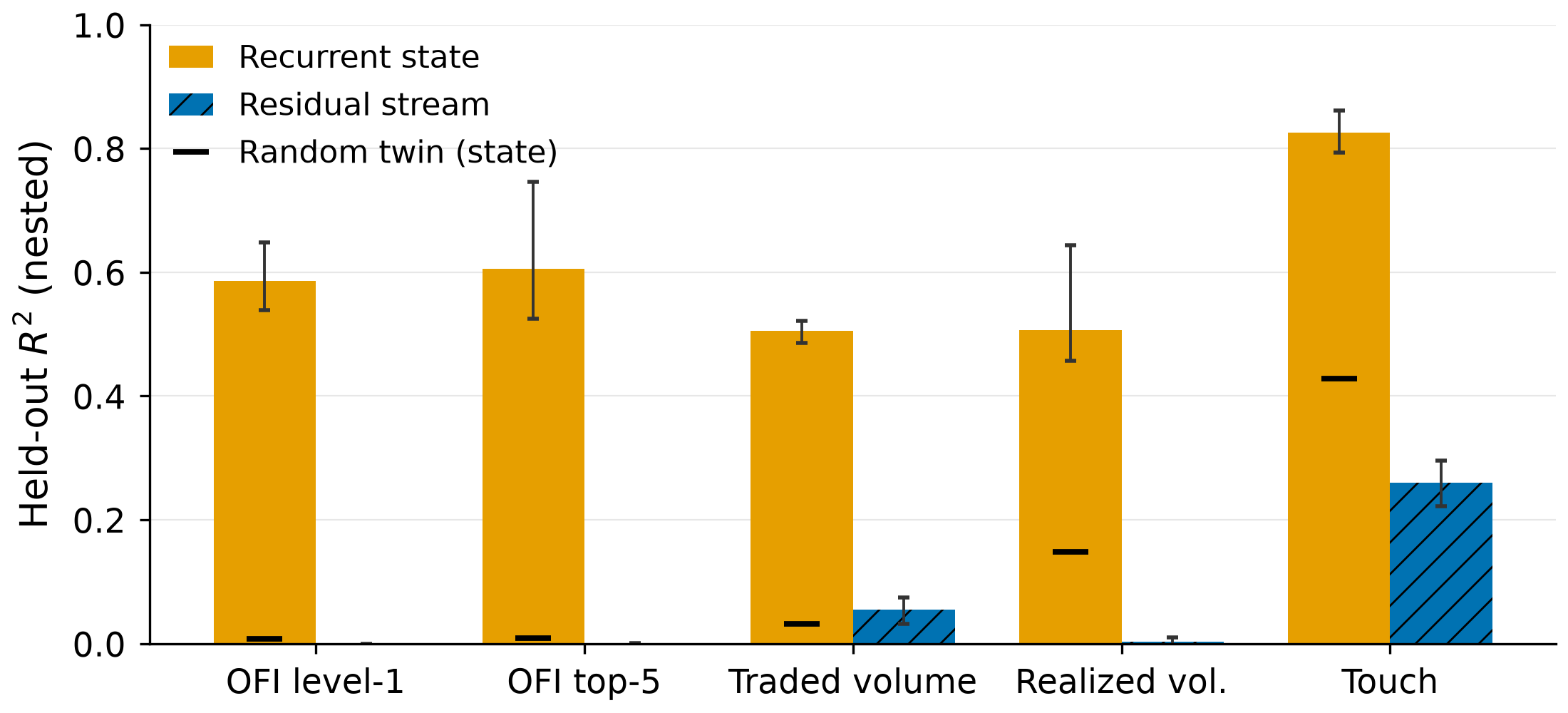}
    \Description{Bar chart: market statistics decode from the recurrent state but not from the residual stream.}
    \caption{Held-out $R^2$ (layer chosen by nested CV) decoded from the recurrent state and the residual stream at $W{=}500$, same windows and probing procedure for both. Whiskers: 95\% day-block bootstrap intervals. OFI, traded volume and realized volatility are decodable almost only from the recurrent state, while the touch is partly readable from both.}
    \label{fig:residual_vs_state}
\end{figure}

In Transformer-based foundation models, mechanistic interpretability probes the residual stream. In our SSM, we find that locus nearly empty of market structure: the representation sits in the recurrent state instead (Figure~\ref{fig:residual_vs_state}). Linear probes on the residual stream recover only a weak readout of the current touch and no integrative market quantities: order-flow imbalance, traded volume, and realized volatility all decode at $R^2 \le 0.06$ from the residual stream at $W{=}500$, against $0.50$--$0.61$ from the recurrent state.

On the other hand, the recurrent state linearly encodes market microstructure. The touch reaches an $R^2$ of $0.83$ in the SSM state compared to only $0.26$ in the residual stream. Comparing $R^2$ performance with a random-weights twin, we find that learning accounts for nearly all of the OFI and traded-volume signal (twin $0.01$ and $0.03$) and raises the touch from $0.43$ to $0.83$. The twin's high touch floor ($0.26$ to $0.43$ across the ten stocks) reflects that recent messages partially echo the current quotes even through random recurrent features. On top of that, a readout of the full book snapshot from about 500 messages earlier reaches only $R^2=0.49$, so the state tracks the current touch better than the book at the start of the window predicts it.

\subsubsection{Order-flow imbalance in the recurrent state}

Our foundation model successfully learns and encodes order-flow imbalance in its recurrent state (Table~\ref{tab:ofi}). We make several observations: First, our model encodes flow, not a snapshot: OFI decodes about five times better than a readout of the full boundary book snapshot ($0.59$ vs.\ $0.12$). Second, unlike touch, the random-weights twin scores $\approx 0$ for OFI at $W{=}500$, so training is the only source of performance there. And lastly, OFI results survive forward-chaining time-series cross-validation at every lookback, with the largest relative margin over the snapshot at $W{=}10$ ($0.26$ vs.\ $0.02$).

\begin{table}[h]
    \centering
    \caption{Held-out OFI $R^2$ (layer chosen by nested CV) from the trained SSM state, compared with a random-weights twin and the best readout, chosen by nested CV, of the full book snapshot at the observation boundary (Book). We include forward-chaining cross-validation for robustness.}
    \label{tab:ofi}
    \Description{Table of held-out R-squared for order-flow imbalance at lookback windows of 500, 100 and 10 messages, comparing the trained state under k-fold and forward-chaining cross-validation with the random-weights twin and the book-snapshot baseline.}
    \begin{tabular}{lcccc}
    \toprule
    Target & Trained & fwd-CV & Random & Book \\
    \midrule
    OFI L1 $W{=}500$       & 0.59 & 0.53 & 0.01 & 0.12 \\
    OFI L1 $W{=}100$       & 0.49 & 0.44 & 0.03 & 0.12 \\
    OFI L1 $W{=}10$        & 0.29 & 0.26 & 0.07 & 0.02 \\
    OFI top-5 $W{=}500$    & 0.61 & 0.51 & 0.01 & 0.08 \\
    \bottomrule
    \end{tabular}
\end{table}

Given the role of OFI as a driver of short-horizon price moves~\cite{cont2014price} and a basis of the deep prediction literature~\cite{kolm2023deep}, the encoding of such flow statistics suggests that LOB state-space foundation models carry the kind of flow information downstream predictors rely on; we test its forward value in Section~\ref{sec:forward}.

\subsubsection{Volume, volatility, and depth}

Decoding is not limited to the touch and OFI: the recurrent state also decodes traded volume, realized volatility, and book depth under both CV protocols (Table~\ref{tab:market_metrics}). Traded volume and realized volatility decode far above both the random-twin floor and the book-snapshot baseline. Depth is the exception: a stale snapshot from about 500 messages earlier predicts boundary depth better ($0.64$) than the state does ($0.38$). Resting liquidity is highly persistent, and a state built from 500 messages has seen only the part of the book those messages touched.

\begin{table}[H]
    \centering
    \caption{Held-out $R^2$ from the SSM state (layer chosen by nested CV) vs.\ the random-weights twin and the best book-snapshot baseline, chosen by nested CV among all three snapshot sets (Book), including forward-chaining CV. For depth, Book is the stale window-start snapshot, since the boundary snapshot contains the target itself.}
    \setlength{\tabcolsep}{4.5pt}
    \Description{Table of held-out R-squared for traded volume, realized volatility and book depth over 500-message windows, comparing the trained state under k-fold and forward-chaining cross-validation with the random-weights twin and the book-snapshot baseline.}
    \begin{tabular}{lcccc}
    \toprule
    Target ($W=500$) & Trained & fwd-CV & Random & Book \\
    \midrule
    Traded volume & 0.50 & 0.44 & 0.03 & 0.11 \\
    Realized volatility & 0.51 & 0.50 & 0.15 & 0.23 \\
    Log-vol L10 book depth & 0.38 & 0.30 & 0.09 & 0.64 \\
    \bottomrule
    \end{tabular}
    \label{tab:market_metrics}
\end{table}

\subsubsection{Per-layer encoding variations}

\begin{figure}[H]
  \centering
  \includegraphics[width=\columnwidth]{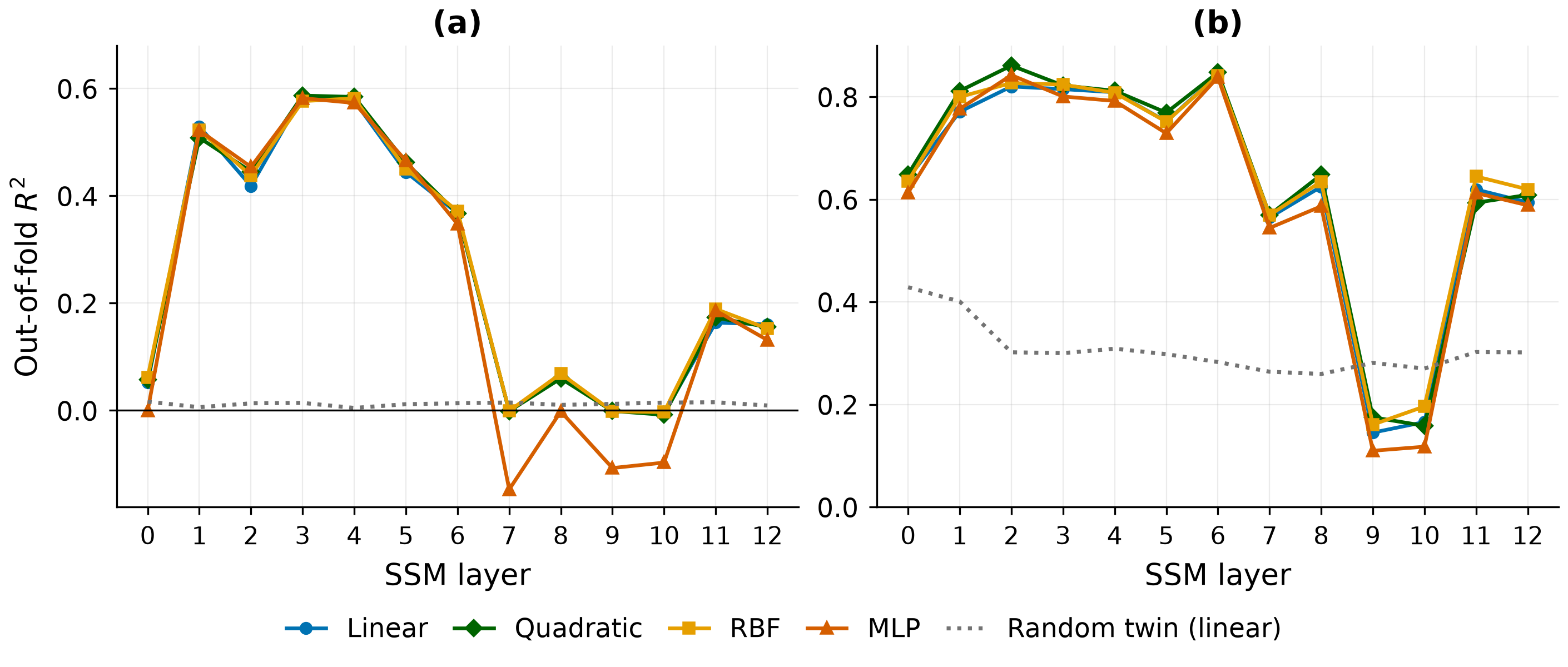}
  \Description{Per-layer R-squared curves for OFI and touch under linear and non-linear probes.}
  \caption{Per-layer decodability for (a)~OFI L1 ($W{=}500$) and (b)~touch (best-ask offset). We test four probes of increasing capacity: linear, quadratic, RBF kernel ridge, and a one-hidden-layer MLP on the top 384 principal components of the state. All four probes show the same pattern, so the mid-stack dip is not a non-linear re-encoding.}
  \label{fig:nonlinear}
\end{figure}

Decodability is far from uniform across the stack (Figure~\ref{fig:nonlinear}).
OFI L1 ($W{=}500$) peaks early in the stack ($R^2=0.59$ at layer 3), falls to near
zero through the middle of the stack, and partially recovers at the penultimate layer
($0.16$); touch follows the same shape at a higher level, recovering to $0.62$. 

The dip is not a non-linear re-encoding. Quadratic, RBF-kernel and one-hidden-layer MLP probes follow the linear curve at every layer, so the mid-stack states do not hold OFI in a form a linear readout misses. At the dip layers the trained state scores very close to or below its random-weights twin.

\subsubsection{Relevance of specific messages}

Probes establish what the state encodes, but not how that code is assembled from the message history. In a message-only model the recurrent state is the entire memory, so the practical question arises: how much history is actually in play, and how long does a single event keep mattering? Mamba-3's state-space duality makes this measurable exactly. The state at the observation boundary is a weighted sum of per-token writes, so summing them within each message attributes it message by message, a recurrent analogue of an attention map, as described in Section~\ref{sec:probing}. We trace two quantities back to the messages that wrote them: the state as a whole, and the single direction the OFI probe reads out.

\begin{figure}[H]
    \centering
    \includegraphics[width=\linewidth]{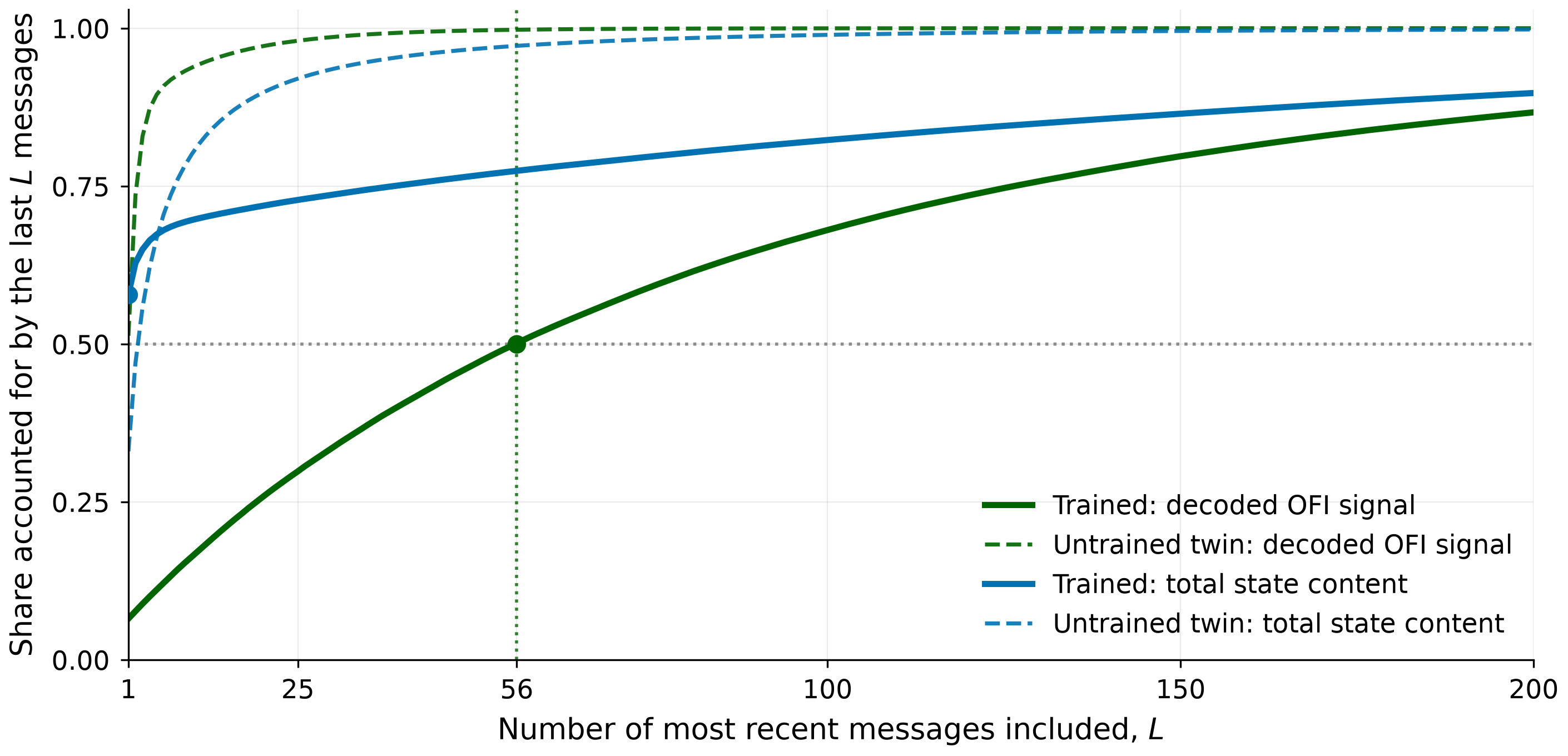}
    \Description{Cumulative share of the state and of the decoded OFI signal against the number of recent messages.}
    \caption{Share of the recurrent state (blue; all layers) and of the decoded OFI signal (green; best layer) accounted for by the $L$ most recent messages, for the trained model and its random-weights twin. Half of the raw state is the newest message alone; half of the OFI signal takes the last 56 messages.}
    \label{fig:attention_lag}
\end{figure}

The two curves come apart (Figure~\ref{fig:attention_lag}). The single newest message accounts for $58\%$ of the total write magnitude, and the last 50 messages for $77\%$: a state snapshot is mostly a description of the present. The decoded OFI signal behaves in the opposite way: the newest message contributes $7\%$, and half the signal needs the last 56 messages. This contrast is between all layers together and the probe layer: at layer 3 itself the state is spread similarly to the OFI signal ($10\%$ from the newest message, half within 61 messages), and the recency mass sits mostly in layer 10. At the probe layer, order flow is held in a long-memory subspace: of its eight heads, four have state half-lives of 28--66 messages while the other four forget within a message. In the random-weights twin both curves collapse onto the newest few messages: $51\%$ of the readout comes from one message, $89\%$ from the last five, and no head at the probe layer exceeds a 10-message half-life. The long-horizon integrators at the probe layer are therefore built by training rather than given by the architecture. 

\subsection{Forward predictions}
\label{sec:forward}

Beyond investigating what the state contains, we test whether the representation can be used: whether interventions on the state move the model's predictions, and whether the state carries signal about the market's near future.  

\subsubsection{Causal steering}
\label{sec:steering}

The most direct readout of a causal change in a model trained autoregressively is through next-token distribution. Following a difference-of-means approach (Section~\ref{sec:steering-method}), we push the recurrent state at the last context message along each encoded feature's axis, simulating high or low flow (OFI), heavy or light trading (traded volume), and turbulent or calm prices (realized volatility), then teacher-force the real future messages, and observe the per-message probability of the pushed property against the same window without a push ($\alpha{=}0$). Doses are expressed in each feature's natural units ($\alpha{=}1$ is one top-vs-bottom tercile displacement) and stay within the axis's seen coordinate range.

\begin{figure*}[t]
    \centering
    \includegraphics[width=\textwidth]{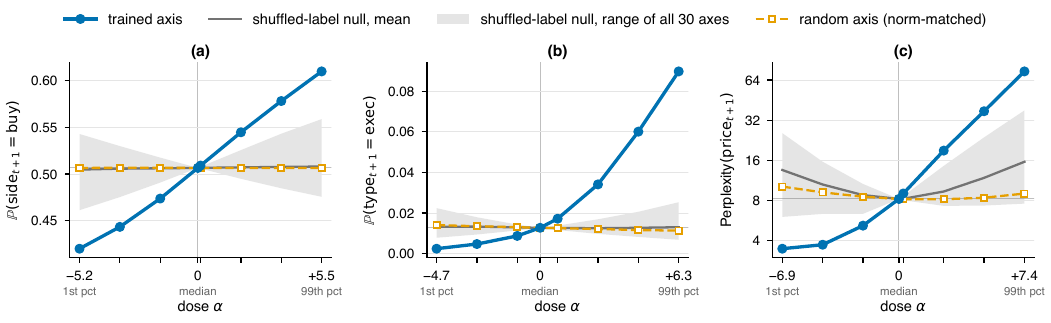}
    \Description{Three side-by-side panels, one per feature, showing the model's next-message belief against steering dose: the trained axis rises monotonically, the shuffled-label null stays near the unpushed value with a band that widens with dose, and the random axis is inert.}
    \caption{Model's beliefs about the next message after pushing each feature's state axis by a dose $\alpha$: (a)~order flow, (b)~traded volume, and (c)~realized volatility. Grey: mean and range of 30 norm-matched axes built from shuffled labels. Orange: a norm-matched random axis. Perplexity measures the effective width in one-tick levels of the next-price forecast.}
    \label{fig:steering_beliefs}
\end{figure*}

The model's beliefs respond cleanly and monotonically for all three features (Figure~\ref{fig:steering_beliefs}). Across the seen range the buy probability moves from $0.42$ to $0.61$; a volume push scales $\mathbb{P}(\mathrm{type}_{t+1}{=}\mathrm{exec})$ by about $1.4\times$ per unit dose, from $0.002$ to $0.090$ end to end; and a turbulence push widens the next-price forecast nine-fold while a calm push sharpens it. At every dose the trained response lies outside the full range of the 30 shuffled-label axes, whose mean stays at the unpushed value for order flow and traded volume. For volatility, the null itself widens the price forecast (higher perplexity) at large doses in either direction, possibly because a push of that size moves the state into a region the model has not seen. In contrast, a calm push along the volatility axis narrows it, although the narrowing levels off at the largest doses, likely in part because it must overcome the generic widening that any large push causes. Additionally, a turbulent push along the same axis widens the forecast far more than any null or random axis. Overall, the responses follow the specific encoded directions, not the size of the push.

Steered beliefs only partially materialize when the model generates messages autoregressively from the steered state and they are replayed through the matching engine. We do so for $50$ generated messages and compare each realized statistic against the shuffled-label null. The volume push carries through to the market: median executed volume rises from $25$ shares at $\alpha{=}0$ to $187$ at $\alpha{=}{+}2.4$ and falls to $1$ at $\alpha{=}{-}2.1$. The flow push carries through to the order flow on the sell side, but not to the price: the generated order flow shifts in the pushed direction for sell pushes, while the mid stays within the null band, with only a weak sell-side drift. The volatility push does not carry through: realized volatility stays within the null band. The model expects prices to disperse, yet does not generate more volatile paths, even though the volume push does raise realized volatility, possibly because its extra executions consume depth. Causal control over the generated market thus ranks traded volume $>$ order flow $>$ volatility, consistent with how directly single messages determine each statistic, which means there is a difference between how decodable and how controllable a feature is (Tables~\ref{tab:ofi} and~\ref{tab:market_metrics}).

\subsubsection{Future-time decoding}

We next ask whether the state frozen at the boundary carries information about the market's near future, without any autoregressive rollout, and whether that information goes beyond a snapshot of the visible book.

A positive held-out $R^2$ indicates a predictive signal. The state carries only a weak one: it explains $4\%$ of the variance of the next-25-message mid-price move and at most $5\%$ of future top-5 OFI, with no signal left beyond $H{\approx}100$. For the mid move, this is level with a probe handed the L10 book snapshot instead of the state ($R^2=0.036$ at $H{=}25$), whereas for future OFI the book snapshot carries almost no signal ($R^2 \le 0.013$). The random-weights twin stays at or below zero at every horizon. The representation thus tracks the current flow but carries little forward information: for the mid move no more than the visible book, and for order flow a small signal the book does not carry.

\subsubsection{Directional accuracy}

\begin{figure}[H]
    \centering
    \includegraphics[width=\linewidth]{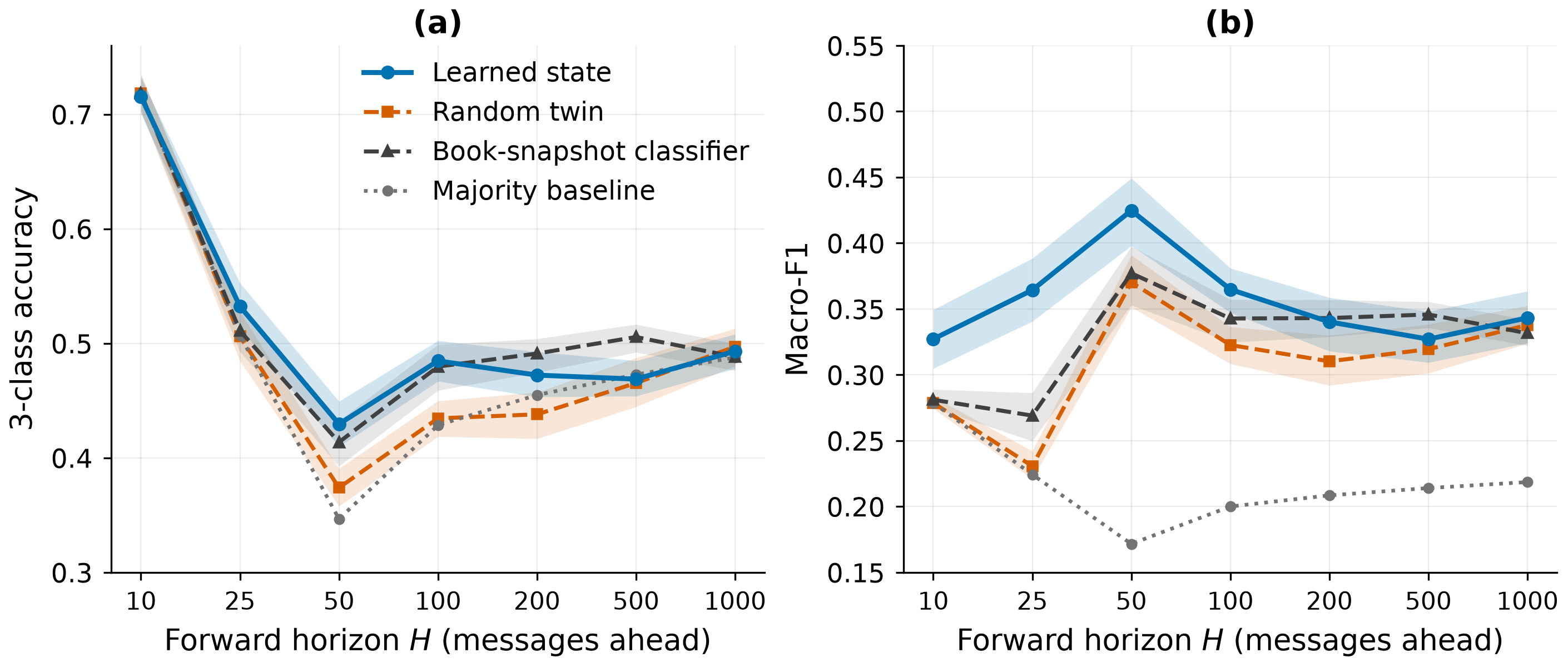}
    \Description{Directional accuracy and macro-F1 against forecast horizon.}
    \caption{Three-class (down/flat/up) mid-price direction from a linear decoding of the recurrent state ($W{=}500$) for different forward horizons $H$: (a)~accuracy, (b)~macro-F1. Grey dashed: book-snapshot baseline classifier, with the best feature set and kernel chosen by nested CV. Shaded bands: 95\% day-block bootstrap CIs.}
    \label{fig:diracc}
\end{figure}

We also test specifically for the direction of the move, separately from its size: we evaluate three-class directional accuracy (down, flat, up) at different horizons (Figure~\ref{fig:diracc}).

Three-class accuracy is high at short horizons but nearly identical for all approaches, including the majority class (Figure~\ref{fig:diracc}a). Accuracy can be misleading here, because most windows at short horizons have no mid-price move, so calling the flat class alone already scores well. Macro-F1, which weights the three classes equally, shows a modest gap: up to $H{=}50$ the state reaches $0.33$, $0.36$ and $0.42$ at $H{=}10$, $25$ and $50$, against $0.28$, $0.27$ and $0.38$ for the book-snapshot baseline and $0.28$, $0.23$ and $0.37$ for the twin. The edge narrows by $H{=}100$, and from $H{=}200$ the book matches or exceeds the state.

We also specifically test for directional accuracy on windows where the mid-price actually moved (i.e.\ binary directional accuracy). In this case, the same probe is right $61.2\%$ of the time at $H{=}25$ and $60.0\%$ at $H{=}50$, against $50.5\%$ for the majority baseline and $50$--$51\%$ for the random-weights twin, but level with the book-snapshot readout ($61.8\%$ and $60.9\%$). Skill fades with horizon, to $57.8\%$ at $H{=}100$ and $53.0\%$ at $H{=}200$, and reaches chance by $H{=}500$.

\subsection{Learning dynamics and cross-stock generalization}
\subsubsection{Pretraining behavior} 
\begin{figure}[h]
    \centering
    \includegraphics[width=\linewidth]{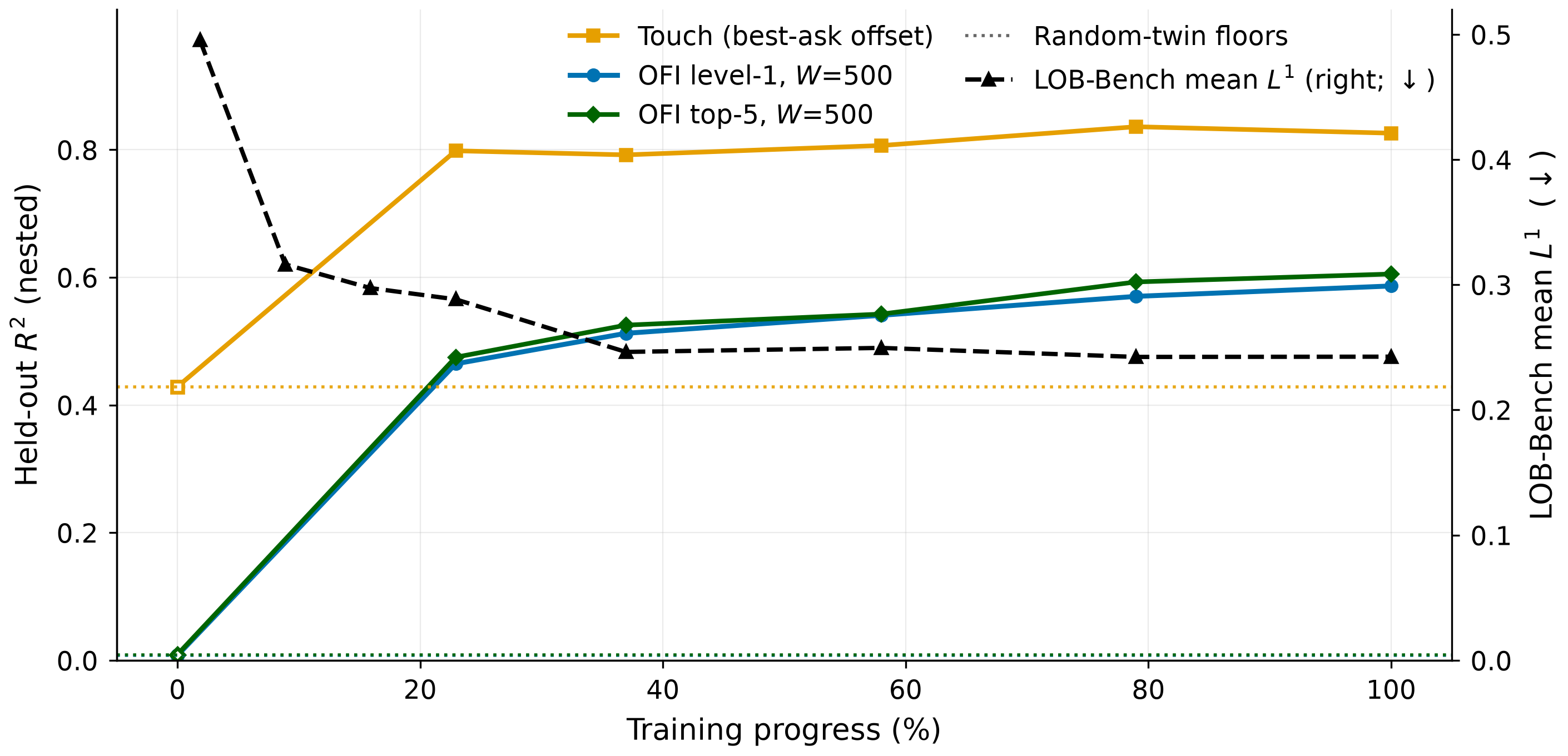}
    \Description{Decodability and LOB-Bench scores across training checkpoints.}
    \caption{Market statistics decodability and LOB-Bench scores at different parts of training (0\% = random-weights twin). Touch is learned early, OFI keeps improving over the whole run, and LOB-Bench plateaus after about a third of training.}
    \label{fig:training_curve}
\end{figure}

Touch and OFI are market features learned early in training (Figure~\ref{fig:training_curve}). Furthermore, while linear decoding of the touch plateaus quickly, OFI's decoding improves steadily throughout training, without a clear plateau. This suggests that despite cross-entropy and LOB-Bench hitting a performance plateau, longer pretraining may reinforce OFI encoding within the recurrent state.

\subsubsection{Generalization across stocks}

\begin{figure}[h]
    \centering
    \includegraphics[width=\linewidth]{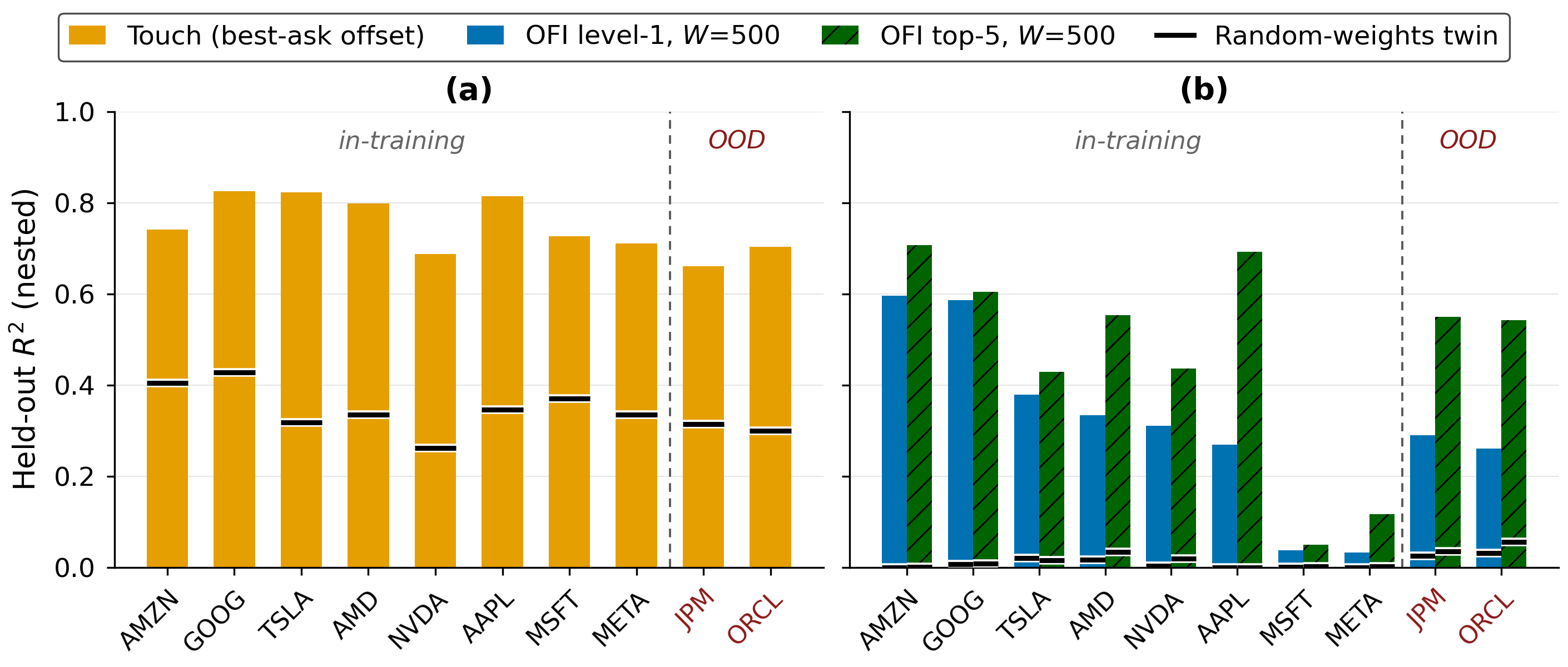}
    \Description{Per-stock R-squared for touch and OFI, including out-of-distribution stocks, with the random-weights twin marked on every bar.}
    \caption{Held-out $R^2$ (layer chosen by nested CV) of probes on the frozen SSM state for ten stocks, $W{=}500$. Black ticks represent the random-weights twin and the dashed divider separates in-training from out-of-distribution (OOD) stocks. (a) Touch decodes on all ten stocks, including the OOD stocks, while (b) OFI $R^2$ varies widely.}
    \label{fig:generalization}
\end{figure}

The recurrent state linearly encodes the touch on all eight training stocks as well as on the two out-of-distribution stocks, which the model never saw in training (Figure~\ref{fig:generalization}). However, raw $R^2$ for realized order-flow imbalance varies across stocks. For level-1 OFI, the state reaches $0.26$ to $0.60$ on all but META and MSFT, where it is near zero on raw $R^2$, whereas a probe reading the boundary book snapshot stays at $R^2 \le 0.14$ and the random-weights twin at $R^2 \le 0.04$ on every one of the ten stocks.

\begin{figure}[h]
    \centering
    \includegraphics[width=\linewidth]{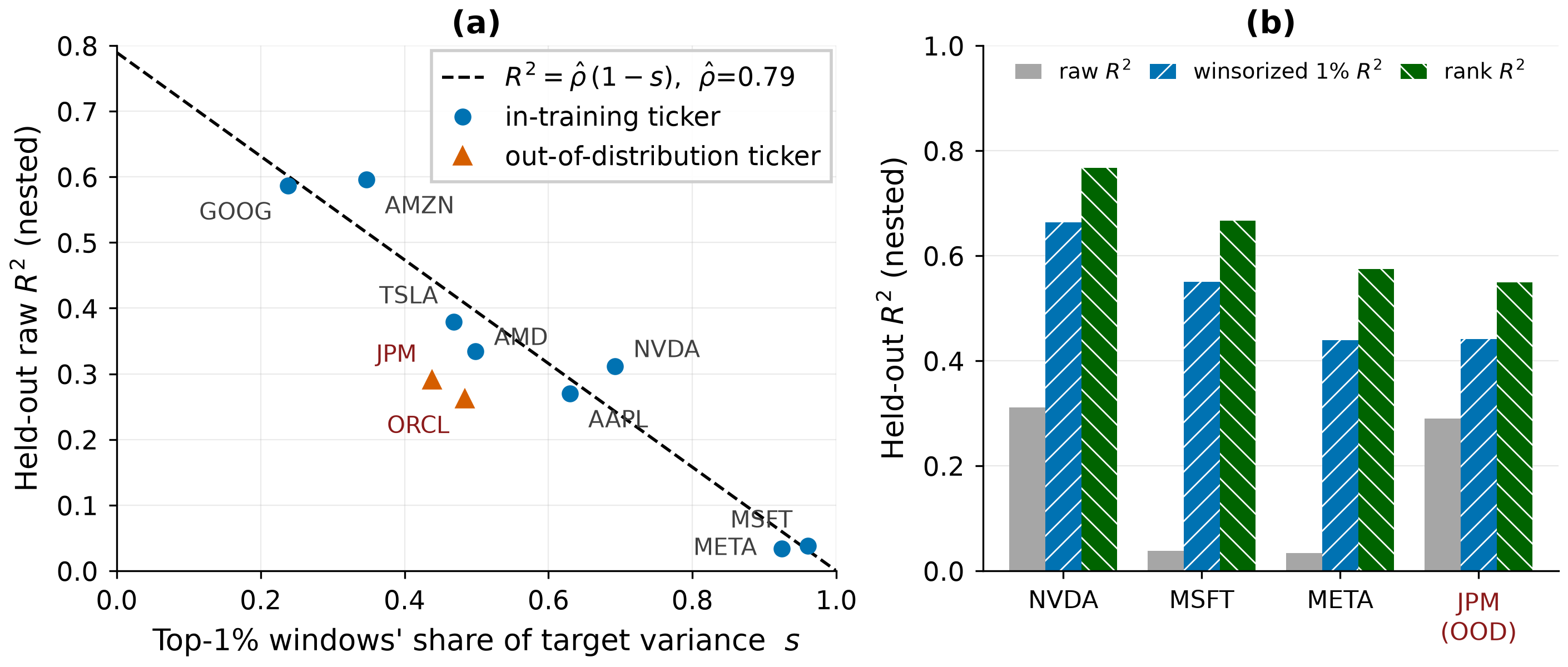}
    \Description{OFI R-squared against the tail share of target variance, and robust-target comparison.}
    \caption{(a)~OFI L1 $R^2$ versus the top-1\% windows' share $s$ of target variance. Tail prevalence explains most of the variation across training stocks (fitted line; fit $R^2 = 0.94$, Spearman $r_s = -0.93$, $p = 0.002$), with the two OOD stocks about 0.15 below. (b)~Raw vs.\ winsorized-1\% vs.\ rank-transform $R^2$ for a representative subset: both tail-robust metrics lift every stock above $R^2 = 0.4$ and narrow the spread across stocks.}
    \label{fig:tail}
\end{figure}

The variation largely tracks how much of each stock's OFI variance sits in rare bursts: out-of-fold $R^2$ decreases linearly with the share $s$ of target variance held by the top-1\% of windows (Figure~\ref{fig:tail}). This relation shows that the linear readout of the state misses extreme bursts: the fitted line is consistent with a probe that explains little of the extreme-burst variance and about 80\% of the remaining variance on the training stocks. To mitigate the effects of tails and recover linear decoding performance, several methods exist. One option is to winsorize OFI, capping the top and bottom 1\% of episodes at the 1st and 99th percentile values. Another option is to use Rank $R^2$, which replaces each episode's OFI value by its normalized rank among all evaluation windows. The resulting $R^2$ rewards recovering the ordering of episodes rather than their magnitude, and is therefore insensitive to heavy tails by construction. Under either metric, OFI is represented across the board for OFI L1 ($R^2 \geq 0.4$) and OFI top-5 ($R^2 \geq 0.5$) for both in-training and out-of-distribution stocks.

\section{Conclusion}

We opened up a message-only foundation model of the limit order book and asked what it learns. Its recurrent state maintains an internal representation of the market's dynamics: order-flow imbalance, traded volume, and realized volatility decode far above random-weights and book-snapshot baselines, the representation remains decodable on stocks never seen in training, and steering it shifts the model's expectations in the direction each pushed feature implies. Trained only to predict the next token of the message stream, the model has learned not only a picture of the market, but its flow.

In future work, unsupervised tools such as sparse autoencoders could be applied to look for features the model has constructed beyond the concepts we knew to probe for, and thereby potentially uncover predictors that the market microstructure literature has not yet identified. Beyond interpretability, the approach also has practical potential. Linear maps read the market dynamics from the state at the cost of a single matrix product, far cheaper than estimating them from autoregressive rollouts, which could be helpful to make foundation models more practical for latency-sensitive applications such as high-frequency trading. Additionally, a steerable state offers a handle for scenario generation without conditioning inputs or retraining. Overall, our work takes a first step toward using LOB foundation models not only as market simulators, but as representation engines whose internal state can be read and steered.

\begin{acks}
    This work was supported by the Engineering and Physical Sciences Research Council [grant number UKRI5149] and other \href{https://www.bold-lab.ai}{sponsors of BOLD}.
    The authors acknowledge the use of resources provided by the Isambard-AI National AI Research Resource (AIRR)~\cite{mcintoshsmith2024isambardaileadershipclasssupercomputer}. Isambard-AI is operated by the University of Bristol and is funded by the UK Government’s Department for Business, Innovation, Science and Trade (DBIST) via UK Research and Innovation; and the Science and Technology Facilities Council [ST/AIRR/I-A-I/1023].
    AB gratefully acknowledges the support of the Anne Dias Griffin and
    Kenneth C. Griffin Harvard University Fellowship. VM gratefully acknowledges funding from the Department of Computer Science at the University of Oxford. AC acknowledges funding from a UKRI AI World Leading Researcher Fellowship (grant EP/W002949/1). AC and JF acknowledge funding from a JPMC Faculty Research Award.
\end{acks}

\bibliographystyle{ACM-Reference-Format}
\bibliography{references}

\end{document}